\documentclass[twocolumn]{esrel2022}

\def\papername{\jobname}

\usepackage{natbib}
\usepackage{tikz}
\usetikzlibrary{automata,arrows,positioning,calc}
\let\captionbox\undefined
\usepackage{caption}
\usepackage{subcaption}
\usepackage{booktabs}
\usepackage{multirow}
\usepackage{soul}
\usepackage{amsmath}
\usepackage{enumerate}
\usepackage{lipsum}
\usepackage[flushleft]{threeparttable}
\usepackage{blindtext}
\usepackage{hyperref}
\usepackage[symbol]{footmisc}

\usepackage{placeins}

\newif\ifsubmit

\definecolor{lightblue}{RGB}{224,224,255}
\definecolor{lightred}{RGB}{255,224,224}
\definecolor{lightgreen}{RGB}{224,255,224}
\definecolor{lightyellow}{RGB}{255,255,224}
\definecolor{lightpurple}{RGB}{255,224,255}
\definecolor{darkerred}{RGB}{64,0,0}
\definecolor{darkred}{RGB}{128,0,0}
\definecolor{darkblue}{RGB}{0,0,128}
\definecolor{darkgreen}{RGB}{0,128,0}
\definecolor{darkpurple}{RGB}{128,0,128}
\newcommand{\colorpar}[3]{\colorbox{#1}{\parbox{#2}{#3}}}
\newcommand{\marginremark}[3]{\marginpar{\colorpar{#2}{\linewidth}{\color{#1}#3}}}

\newcommand{\zenodo}{\href{https://zenodo.org/record/6535853}{\mbox{zenodo.org/record/6535853}}}

\newcommand{\foottt}{All comparative figures and scripts are available at }

\ifsubmit
    \newcommand{\LJ}[1]{}
    \newcommand{\MS}[1]{}
    \newcommand{\TT}[1]{}
    \newcommand{\HM}[1]{}
    \newcommand{\TH}[1]{}
\else
    \newcommand{\LJ}[1]{\marginremark{darkblue}{lightblue}{\scriptsize{[LJ]~ #1}}}
    \newcommand{\MS}[1]{\marginremark{darkpurple}{lightpurple}{\scriptsize{[MS:]~#1}}}
    \newcommand{\HM}[1]{\marginremark{darkred}{lightred}{\scriptsize{[HM]~#1}}}
    \newcommand{\TT}[1]{\marginremark{darkred}{lightred}{\scriptsize{[TT]~#1}}}
    
\fi

\begin{document}

\markboth{L.A. Jimenez-Roa, T. Heskes, T. Tinga, H. Molegraaf and M. Stoelinga}{Deterioration modeling of sewer pipes via discrete-time Markov chains}

\twocolumn[

\title{Deterioration modeling of sewer pipes via discrete-time Markov chains: A large-scale case study in the Netherlands}

\vspace{-17pt}

\author{Lisandro A. Jimenez-Roa}

\address{Formal Methods and Tools (FMT), University of Twente, The Netherlands. \email{l.jimenezroa@utwente.nl}}

\author{Tom Heskes}

\address{Institute for Computing and Information Sciences, Radboud University Nijmegen, The Netherlands. \email{Tom.Heskes@ru.nl}}

\author{Tiedo Tinga}

\address{Dynamics Based Maintenance (DBM), University of Twente, The Netherlands. \email{t.tinga@utwente.nl}}

\author{Hajo J. A. Molegraaf}

\address{Rolsch Assetmanagement, The Netherlands. \email{hajo.molegraaf@rolsch.nl}}

\author{Mari\"{e}lle Stoelinga}
\address{Formal Methods and Tools (FMT), University of Twente, The Netherlands. \email{m.i.a.stoelinga@utwente.nl}}

\begin{abstract} 

Sewer pipe network systems are an important part of civil infrastructure, and in order to find a good trade-off between maintenance costs and system performance, reliable sewer pipe degradation models are essential. In this paper, we present a large-scale case study in the city of Breda in the Netherlands.
Our dataset has information on sewer pipes built since the 1920s and contains information on different covariates. We also have several types of damage, but we focus our attention on infiltrations, surface damage, and cracks. Each damage has an associated severity index ranging from 1 to 5. To account for the characteristics of sewer pipes, we defined 6 cohorts of interest.
Two types of discrete-time Markov chains (DTMC), which we called Chain `Multi' and `Single' (where Chain `Multi'contains additional transitions compared to Chain `Single'), are commonly used to model sewer pipe degradation at the pipeline level, and we want to evaluate which suits better our case study. To calibrate the DTMCs, we define an optimization process using Sequential Least-Squares Programming to find the DTMC parameter that best minimizes the root mean weighted square error.
Our results show that for our case study there is no substantial difference between Chain `Multi' and `Single', but the latter has fewer parameters and can be easily trained. Our DTMCs are useful to compare the cohorts via the expected values, e.g., concrete pipes carrying mixed and waste content reach severe levels of surface damage more quickly compared to concrete pipes carrying rainwater, which is a phenomenon typically identified in practice.

\end{abstract}


]




\section{Introduction}\label{sec:introduction}

Sewer network systems are an important part of the civil infrastructure required to achieve an adequate level of social and economic welfare.
The management of these systems has become increasingly challenging due to the need to cope with limited budgets, environmental changes, uncertainty about network deterioration, and a lack of rigorous degradation analysis. This often leads to conservative approaches that result in the early replacement of sewer pipes.

Thus, aiming at finding a good trade-off between maintenance costs and system performance, robust and reliable \textit{sewer pipe degradation models} are needed to prioritize pipes at high risk of failure for proactive maintenance, support decision making, and strategic rehabilitation planning \citep{scheidegger2011network,egger2013sewer}. 

Moreover, there is a need in the research community for sharing existing case studies aiming at increasing the evidence on sewer pipe degradation models \citep{tscheikner2019sewer}.

Concerning sewer pipe deterioration models, three types can be identified: those based on physics, artificial intelligence, and statistics. 
A detailed review of the different types of models used to predict the degradation of sewer networks is presented in \cite{hawari2020state}.

Physics-based models may be too complex to capture the complete degradation behavior, and artificial intelligence models require high computational costs and demands of data \citep{ana2010modeling}. Thus, given the limitations of these types of models, we center our attention on statistical methods. 

In particular, we are interested in Markov chain models, since they proved to be among the most reliable and widely used approaches to simulate sewer pipe deterioration \citep{kobayashi2012statistical,tscheikner2019sewer}, and enable the modeling of sequential events, such as sewer pipes deterioration \citep{ana2010modeling}.

Several types of Markov chains (MC) have been implemented for the modeling of sewer pipe networks, examples are discrete-time MC \citep{micevski2002markov,baik2006estimating}, continuous-time MC \citep{lin2019integrative}, non-Homogeneous MC \citep{le2008modelling}, hidden-MC \citep{kobayashi2012statistical}, semi-MC \citep{scheidegger2011network}. 

As a first step, we decided to use discrete-time Markov chains (DTMCs) because these proved to be a straightforward approach to model degradation patterns associated with sewer pipe networks. Moreover, we are interested in two typical types of DTMCs (see Fig. \ref{fig:markov_chains}) that we call Chains \textit{`Multi'} and \textit{`Single'}, where the former contains additional transitions compared to the latter, and we are interested in evaluating which of them best suits our case study.

Similar to \cite{caradot2018practical}, our goal with these DTMCs is to predict the probability for a pipe to be in a severity class for a certain type of damage, based on the pipe age and a set of numerical or categorical variables (called covariates) organized in 6 cohorts (i.e., group of pipes with the same characteristics) of interest. 

Our research questions are both application-oriented (RQ1) and methodological (RQ2): \textbf{RQ1} how do the predefined cohorts compare in terms of deterioration rate? \textbf{RQ2} how can DTMCs assist in getting this insight, and how do Chains `Multi' and `Single' compare in terms of performance? 

The experimental evaluation is based on a large-scale case study in the city of Breda in the Netherlands, where we have information on sewer pipes built since the 1920s which contains information on different covariates. 
We focus on three typical types of sewer pipes damages namely \textit{infiltration}, \textit{surface damage}, and \textit{cracks}. Each damage has an associated severity index ranging from 1 to 5.

Our main contribution is to demonstrate the application of existing degradation models in a large-scale case study. The present work is a valuable step toward the development of an evidence-based asset management framework. The scripts and comparative figures can be found at \zenodo.

The structure of this paper is as follows. Section \ref{sec:background_DTMC} provides the theoretical background on DTMCs. Section \ref{sec:methodology} presents our methodology. In Section \ref{sec:exp_evaluation} we preset the case study, the experimental evaluation and the main results. We discuss and conclude in Section \ref{sec:discussion_conclusion}.

\section{\mbox{Homogeneous discrete-time Markov chain}}\label{sec:background_DTMC}

A Markov chain is a stochastic process used to describe the deterioration of sewer pipes through condition states \citep{hawari2020state}. 
Among the different types of Markov chains, we adopt a discrete-time Markov chain (DTMC) because it is the simplest type suitable for modeling the degradation of sewer pipes \citep{micevski2002markov}. 

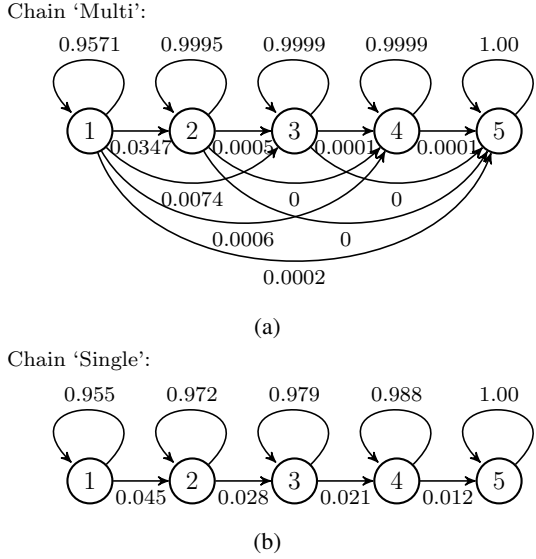
\begin{figure}[!h]
\begin{subfigure}[b]{0.49\textwidth}
\begin{tikzpicture}[->, >=stealth', auto, semithick, node distance=2cm,font=\sffamily\footnotesize]
\node[text width=2.2cm] at (0,1.6) {Chain `Multi':};
\tikzstyle{every state}=[fill=white,draw=black,thick,text=black,scale=0.68,font = {\Large}]
\node[state]    (A)       {$1$};
\node[state]    (B)[right of=A]   {$2$};
\node[state]    (C)[right of=B]   {$3$};
\node[state]    (D)[right of=C]   {$4$};
\node[state]    (E)[right of=D]   {$5$};
\path
(A) edge[loop,above]                node{$0.9571$}         (A)
	edge[below]                     node{$0.0347$}         (B)
	edge[bend right=45,below]    node{$0.0074$}         (C)
	edge[bend right=60,below]    node{$0.0006$}         (D)
	edge[bend right=70,below]    node{$0.0002$}         (E)
(B) edge[loop,above]                node{$0.9995$}         (B)
	edge[below]                      node{$0.0005$}         (C)
	edge[bend right=45,below]          node{$0$}         (D)
	edge[bend right=60,below]          node{$0$}         (E)
(C) edge[loop,above]               node{$0.9999$}         (C)
	edge[below]                      node{$0.0001$}         (D)
    edge[bend right=45,below]          node{$0$}         (E)
(D) edge[loop,above]                node{$0.9999$}         (D)
	edge[below]                      node{$0.0001$}         (E)
(E) edge[loop,above]                node{$1.00$}         (E);
\end{tikzpicture}
\caption{}
\end{subfigure}
\begin{subfigure}[b]{0.49\textwidth}
\begin{tikzpicture}[->, >=stealth', auto, semithick, node distance=2cm,font=\sffamily\footnotesize]
\node[text width=2.2cm] at (0,1.6) {Chain `Single':};
\tikzstyle{every state}=[fill=white,draw=black,thick,text=black,scale=0.68,font = {\Large}]
\node[state]    (A)       {$1$};
\node[state]    (B)[right of=A]   {$2$};
\node[state]    (C)[right of=B]   {$3$};
\node[state]    (D)[right of=C]   {$4$};
\node[state]    (E)[right of=D]   {$5$};
\path
(A) edge[loop,above]               node{$0.955$}         (A)
	edge[below]               node{$0.045$}         (B)
(B) edge[loop,above]                node{$0.972$}         (B)
	edge[below]               node{$0.028$}         (C)
(C) edge[loop,above]                node{$0.979$}         (C)
	edge[below]               node{$0.021$}         (D)
(D) edge[loop,above]                node{$0.988$}         (D)
	edge[below]               node{$0.012$}         (E)
(E) edge[loop,above]                node{$1.00$}         (E);
\end{tikzpicture}
\caption{}
\end{subfigure}
\caption{Example of DTMC modeling the degradation of sewer pipes considering five degradation states (i.e., $K=5$) and no repairs. (a) Chain `Multi'; (b) Chain `Single'.}\label{fig:markov_chains}
\vspace{-8pt}
\end{figure}

A DTMC is a directed graph whose nodes are called states, and whose edges are called transitions. In our case, the states are the possible sewer pipe damage severity (see Fig. \ref{fig:markov_chains}). When the transition probabilities remain constant over time, we talk about \textit{homogeneous Markov chains}.
The \textit{time interval} is a discrete time period (e.g., one year) representing a single transition (or step $n$).

From \cite{baik2006estimating}, a Markov chain is a discrete-time stochastic process, where the probability of any future event depends only on the present state and is independent of the past states. The latter is known as the \textit{Markov property} and can be formally expressed in Eq. \ref{eq:markov_property} for the states $i_0,i_1,\hdots,i_n,i_{n+1}$ and all $n\geq0$ as, 
\vspace{-10pt}
\begin{multline}
P(X_{n+1} = i_{n+1} | X_n = i_n, X_{n-1}=i_{n-1}, \\
\hdots, X_0=i_0) \\
= P(X_{n+1} = i_{n+1} | X_n = i_n),\label{eq:markov_property}
\end{multline} 

\noindent where $X_n$ is the state of a sewer pipe at step $n$. The DTMC assumes that the conditional probability does not change over time i.e., for all the states $i$ and $j$ and all $n$, ${P(X_{n+1}=j | X_{n}=i)}$ is independent of $n$,
\vspace{-2pt}
\begin{equation}
p_{ij} = P(X_{n+1} = j | X_{n} = i), \label{eq:pij}
\end{equation}

\noindent where $p_{ij}$ is the \textit{transition probability} that given a system in state $i$ at step $n$, it will be in a state $j$ at step $n+1$. Since $p_{ij}$ is a probability, it is bounded in the closed interval $0\leq p_{ij} \leq 1$. The transition probabilities are often expressed in a $K \times K$ \textit{transition probability matrix} $P$, where $K$ is the total number of states in the DTMC,
\vspace{-2pt}
\begin{equation}
P =
\begin{bmatrix}
p_{11} & p_{12} & p_{13} & \hdots & p_{1K} \\
p_{21} & p_{22} & p_{23} & \hdots & p_{2K}  \\
p_{31} & p_{32} & p_{33} & \hdots & p_{3K}  \\
\vdots & \vdots & \vdots & \ddots & \vdots  \\
p_{K1} & p_{K2} & p_{K3} & \hdots & p_{KK}
\end{bmatrix}\label{eq:transition_probability_matrix}
\end{equation}

Here $\sum_{j}^K p_{ij} = 1$ for all $ i = 1,2, \hdots , K$. The values $p_{i,j}$ in Fig. \ref{fig:markov_chains} are discussed in Section \ref{sec:comparing_chains}. The \textit{initial state vector} $\bar{S}^{(0)}$ indicates the probability that the system is in the state $k$ at the step $n=0$ (where $1 \leq k \leq K, k \in \mathbb{N}$),
\vspace{-1pt}
\begin{equation}
\bar{S}^{(0)} =
\begin{bmatrix}
S_{1}^{(0)},
S_{2}^{(0)},
S_{3}^{(0)},
\hdots,
S_{k}^{(0)}
\end{bmatrix}^T,\label{eq:S0}
\end{equation}

\noindent where $\sum_{l=1}^k S_l^{(n)} = 1$ at any step $n$. In order to calculate the state probabilities associated to the $n$-step $S^{(n)}$, we apply the Chapman-Kolmogorov equation \citep{stewart2009probability},
\vspace{-2pt}
\begin{equation}
    \begin{aligned}
        \bar{S}^{(n)} &= \bar{S}^{(0)}P^{(n)}\\
        &=
        \begin{bmatrix}
        S_{1}^{(n)},
        S_{2}^{(n)},
        S_{3}^{(n)},
        \hdots,
        S_{k}^{(n)}
        \end{bmatrix}^T \label{eq:chapman-kolmogorov}
    \end{aligned}
\end{equation}

\noindent In Eq. \ref{eq:chapman-kolmogorov}, the $n$-step transition probability matrix $P^{(n)}$ is obtained by multiplying the matrix $P$ by itself $n$ times, i.e., $P^{(n)} = P^n$. 

We can compute the expected severity $E^n$ as done in \cite{baik2006estimating}, by multiplying the state probability vector $\bar{S}^{(n)}$ at a step $n$ by the severity class vector $\bar{C}=[1,2,3,4,5]$. Here $1$ indicates pristine condition and $5$ is the maximal severity that can be possibly assigned to a type of damage.
\vspace{-12pt}
\begin{equation}
    E^n= \bar{S}^{(0)}P^{(n)}\bar{C}^T\label{eq:expectation}
\end{equation}

We adopt two typical chains to model the degradation of sewer pipes via DTMCs \citep{ana2010modeling} (Fig. \ref{fig:markov_chains}), where:
\begin{enumerate}[i]
    \item the states $k$ are associated to the damage severities i.e., the classes in  $\bar{C}$;
    \item for both chains, the transitions can only occur from the current state to worse-case states (i.e., $p_{ij} = 0$ for $i>j$), since it is impossible for a pipe to improve its condition without interventions (i.e., repairs);
    \item only the last state in both chains is absorbing, that is $p_{55} = 1.00$;
    \item the Chain `Multi' (Fig. \ref{fig:markov_chains}.a) allows transitions between consecutive and non-consecutive states (i.e., $0 \leq p_{ij} \leq 1$ for all $i \leq j$) \citep{micevski2002markov,baik2006estimating,scheidegger2011network};
    \item the Chain `Single' (Fig. \ref{fig:markov_chains}.b) is a simplified version of Chain `Multi', and allows transitions only between consecutive states (i.e., $0 \leq p_{ij} \leq 1$ for all $i$ and $ j = i+1$, and $p_{ij} = 0$ for $j>i+1$) \citep{le2008modelling,scheidegger2011network,lin2019integrative}.
\end{enumerate}

\section{Methodology}\label{sec:methodology}

Our goal is to describe the degradation of sewer pipes based on a historic set of inspection data. To achieve this we calibrate discrete-time Markov chains (DTMCs) that quantify the probability of a pipe (from the historic data set) being in a condition class given the age of the pipe. These DTMCs represent cohorts of pipes, i.e., they are trained with data from pipes that share similar characteristics. An overview of the four steps we follow are:
\vspace{-4pt}
\begin{enumerate}[(i)]
    \item pre-processing the data (cleaning);
    \item definition of cohorts; 
    \item creating a \textit{discretized table} per cohort. This is the input data to calibrate the DTMC;
    \item calibration of the DTMC;
\end{enumerate}
\vspace{-4pt}
The details about each of the steps are provided in the following sections.
\vspace{-10pt}
\begin{table}[!h]
\centering
\small
\caption{Cohorts of interest, Fraction (\%) of total number of inspected pipes (25'507).}\label{tb:cohorts}
\begin{tabular}{@{}cll@{}}
\toprule
\multicolumn{1}{c}{\begin{tabular}[c]{@{}c@{}}Cohort\\ name\end{tabular}} & \multicolumn{1}{c}{Description} & \multicolumn{1}{c}{\begin{tabular}[c]{@{}c@{}}Fraction\\ (\%)\end{tabular}} \\ \midrule
CMW & \begin{tabular}[c]{@{}l@{}}Material: Concrete \& \\ Content: Mixed and Waste\end{tabular} & 59.29 \\
CR & \begin{tabular}[c]{@{}l@{}}Material: Concrete \& \\ Content: Rainwater\end{tabular} & 12.85 \\
PMW & \begin{tabular}[c]{@{}l@{}}Material: PVC \& \\ Content: Mixed and Waste\end{tabular} & 18.88 \\
PR & \begin{tabular}[c]{@{}l@{}}Material: PVC \& \\ Content: Rainwater\end{tabular} & 7.89 \\
CdL & \begin{tabular}[c]{@{}l@{}}Material: Concrete \& \\ Width $<$ 500 mm\end{tabular} & 50.16\\
CdG & \begin{tabular}[c]{@{}l@{}}Material: Concrete \& \\ Width $\geq$ 500 mm\end{tabular} & 22.02 \\ \bottomrule
\end{tabular}
\end{table}
\vspace{-20pt}
\subsection{data pre-processing}\label{sec:preprocessing}

Our work is based on the case study described in Section \ref{sec:case_study}. For each sewer pipe the data set contains information about inspections carried out over the years, including (i) inspection's unique identifiers (ii) inspection date, (iii) damage size, (iv) damage codes (unique identifier to a type of damage), (v) damage class (this is what the severity vector $\bar{C}$ in Eq. \ref{eq:expectation} describes), and (vi) relative position of the damage. 

It is worth mentioning that each damage code has an associated damage class. We ignore data associated with pipes built before 1920 and those that have missing/erroneous construction year. 

\begin{table*}[!h]
\centering
\small
\caption{Discretized table $\hat{S}_{k}^{(\hat{n})}$ for cohort CMW, surface damage (BAF), with $\Delta t = 3 \text{ years}$.}
\begin{tabular}{@{}ccccccccc@{}}
\toprule
\multirow{2}{*}{\begin{tabular}[c]{@{}c@{}}Count\\ ($c$)\end{tabular}} &
\multirow{2}{*}{\begin{tabular}[c]{@{}c@{}}PipeAge \\ (years)\end{tabular}} &
\multirow{2}{*}{\begin{tabular}[c]{@{}c@{}}Time \\ ($t$)\end{tabular}} &
\multirow{2}{*}{\begin{tabular}[c]{@{}c@{}}Step\\ ($\hat{n}$)\end{tabular}} & \multicolumn{5}{c}{$\hat{S}_{k}^{(\hat{n})}$} \\ \cmidrule(l){5-9} 
 &  & &  & $k=1$ & $k=2$ & $k=3$ & $k=4$ & $k=5$ \\ \midrule
832 & {[}0,3{)} & 1.5 &  0 & 0.95 & 0.03 & 0.01 & 0.01 & 0.00 \\
$\vdots$ & $\vdots$ &  $\vdots$ & $\vdots$ & $\vdots$ & $\vdots$ & $\vdots$ & $\vdots$ & $\vdots$ \\
2'339 & {[}48,51{)} & 49.5 & 16 & 0.35 & 0.50 & 0.12 & 0.02 & 0.01 \\
$\vdots$ & $\vdots$ & $\vdots$ & $\vdots$ & $\vdots$ & $\vdots$ & $\vdots$ & $\vdots$ & $\vdots$ \\
64 &  {[}75,78{)} & 76.5 & 25 & 0.44 & 0.20 & 0.28 & 0.05 & 0.03 \\
$\vdots$ & $\vdots$ & $\vdots$ & $\vdots$ & $\vdots$ & $\vdots$ & $\vdots$ & $\vdots$ & $\vdots$ \\ \bottomrule
\end{tabular}\label{tb:discretized_table}
\end{table*}

\subsection{definition of cohorts}

In order to account for explanatory variables (other than pipe age) in deterioration modeling, it is necessary to construct cohorts (i.e., groups of sewer pipes that share similar characteristics) and calibrate a DTMC per each cohort.

Table \ref{tb:cohorts} presents 6 cohorts of interest, and the number of pipes with certain characteristics (as a fraction of the total number of inspected pipes). We note that a drawback of defining cohorts is that it could result in small subsets (e.g., Cohort PR), which might not be statistically representative.

\subsection{discretized table}

We build a \textit{discretized table} per cohort (Table \ref{tb:discretized_table}) and damage code, this is the input to calibrate the DTMCs. For this, we identify the \textit{state of a sewer pipe} as the \textit{maximum damage class found during an inspection} for damage codes of interest. This approach is conservative and is important in determining which pipes should be repaired in the near future.

To build Table \ref{tb:discretized_table}, we define a time interval $\Delta t$, and make groups of pipes by age at the time of inspection, then count, per group, how many pipes were found in each damage class and normalize with respect the total number of pipes in the group.

For example, in Table \ref{tb:discretized_table}, for the Cohort CMW, damage code \textit{BAF} (surface damage), and $\Delta t = 3 \text{ years}$, there were $2'339$ pipes with $48 \leq \text{PipeAge} < 51$ years at the time of inspection. Here the \textit{count} vector ($c$) indicates the total number of pipes found within a PipeAge interval.

The time $t$ is the time calculated as the mean value of the PipeAge interval, and $\hat{n}$ is the step associated to the discretization. Thus, the step $\hat{n} = 16$ is associated to the time interval $48 \leq \text{PipeAge} < 51$ at $t=49.5$ years.

In this interval, 35\% of the pipes were in State 1 \big(i.e., $\hat{S}_{k=1}^{(\hat{n} = 16)} = 0.35$\big), 50\% in State 2 \big(i.e., $\hat{S}_{k=2}^{(\hat{n} = 16)} = 0.50$\big), and so on. Therefore, $\hat{S}_k^{(\hat{n})}$ is a $|\hat{n}| \times K$ matrix and represents the \textit{ground truth used to calibrate the DTMCs}.

The sum of the counts ($c$) yields the total number of pipes in the network. Notice that $c$ varies in each PipeAge interval. We consider this when calibrating the DTMCs by defining a \textit{weight} vector.

It is worth mentioning that we ignore the \textit{right-censoring} in our data set, this means that we assume that the sewer pipe has just moved to the condition as observed during the inspections, which may not be the case, as the pipe could have entered that condition prior to the inspection.

\subsection{calibration of the DTMC}\label{sec:DTMC_parameter_inference}

To calibrate a DTMC we implement an optimization process where the objective is to find the set of parameters in the DTMC, namely $\bar{S}^{(0)}$ (Eq. \ref{eq:S0}) and $P$ (Eq. \ref{eq:transition_probability_matrix}), that minimize the \textit{Root Mean Weighted Square Error} ($Err$) (Eq. \ref{eq:error_measure}).
For this we first compute a weight vector $\bar{w}$, which results from the normalization of the counts ($\bar{c}$) (Table \ref{tb:discretized_table}) with respect its maximum value,
\vspace{-4pt}
\begin{equation}
\bar{w} = \frac{c}{\max(c)}, \label{eq:weight}
\end{equation}

\noindent $Err$ is computed as the difference between the discretized table \big($\hat{S}^{(\hat{n})}_{k}$\big), and the predictions made with the DTMC for the same steps $\hat{n}$ using Eq. \ref{eq:chapman-kolmogorov} \big($\bar{S}^{(\hat{n})}_{k}$\big),
\vspace{-10pt}
\begin{equation}
Err = \sqrt{ \frac{ \sum_{\hat{n},k} \big[ \big(\bar{S}^{(\hat{n})}_{k} - \hat{S}^{(\hat{n})}_{k}\big)^2 * \bar{w}_{\hat{n}}\big] }{|\hat{n}| \times K} }\label{eq:error_measure}
\end{equation}

\vspace{-4pt}
The minimization of $Err$ is carried out through the \textit{Sequential Least-Squares Programming} (SLSQP) algorithm available in Scipy \citep{2020SciPy-NMeth}, and we use the default parameters. All the optimization parameters are bounded in the closed interval $[0,1]$, and are always initialized in the same form: (i) in $\bar{S}^{(0)}$, $\bar{S}_{k=1}^{(0)}=1$ and $\bar{S}_{k\neq1}^{(0)}=0$, (ii) $P$ is the identity matrix. We adopt the constraints described at the end of Section \ref{sec:background_DTMC} for both Markov chains.

We calibrate the DTMCs by randomly selecting 50\% of the available data per cohort using repeated half-sample bootstrap \citep{saigo2001repeated}. After convergence, the output of the calibration process are the $\bar{S}^{(0)}$ and $P$ with the smallest error $Err$ for a given $\Delta t$.


\section{Experimental Evaluation}\label{sec:exp_evaluation}

\subsection{Case study}\label{sec:case_study}
Our case study consists of a large-scale sewer pipe network in the city of Breda. The network is composed of 25'723 (1'052 km) sewer pipes, mostly built from 1950 onwards.

Most of the pipes are made out of concrete (72\%) and PVC (27\%), have rounded (94\%) and ovoid (5.4\%) shapes, are used for transport (98\%), are less than 170 meters long (99.9\%), have a diameter up to 1 meter (98.3\%), have different content such as mixed (63\%), rainwater (21\%), and waste (16\%). 

\begin{figure*}[!h]
\centering
\begin{subfigure}[b]{0.245\textwidth}
 \centering
 \includegraphics[width=\textwidth]{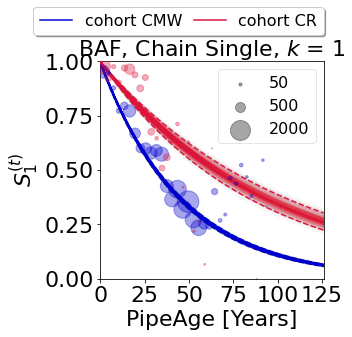}
\caption{}
\end{subfigure}
\begin{subfigure}[b]{0.245\textwidth}
 \centering
 \includegraphics[width=\textwidth]{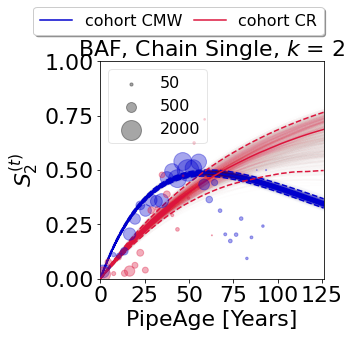}
\caption{}
\end{subfigure}
\begin{subfigure}[b]{0.245\textwidth}
 \centering
 \includegraphics[width=\textwidth]{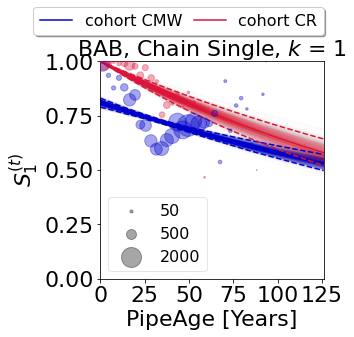}
\caption{}
\end{subfigure}
\begin{subfigure}[b]{0.245\textwidth}
 \centering
 \includegraphics[width=\textwidth]{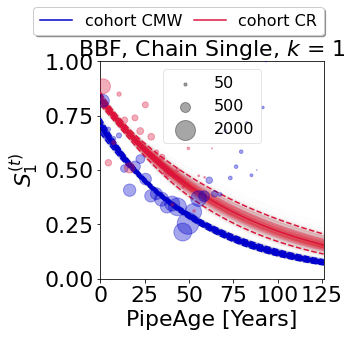}
\caption{}
\end{subfigure}
\begin{subfigure}[b]{0.235\textwidth}
     \centering
     \includegraphics[width=\textwidth]{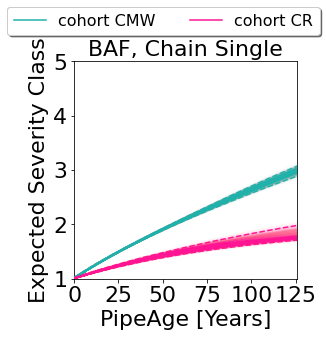}
\caption{}
\end{subfigure}
\begin{subfigure}[b]{0.235\textwidth}
     \centering
     \includegraphics[width=\textwidth]{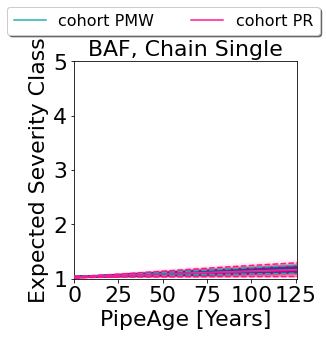}
\caption{}
\end{subfigure}
\begin{subfigure}[b]{0.235\textwidth}
     \centering
     \includegraphics[width=\textwidth]{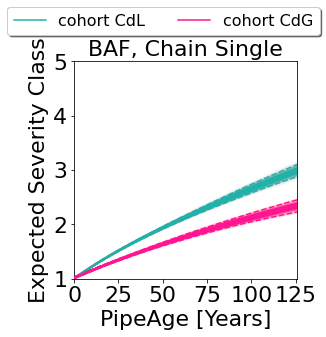}
\caption{}
\end{subfigure}
\begin{subfigure}[b]{0.247\textwidth}
     \centering
     \includegraphics[width=\textwidth]{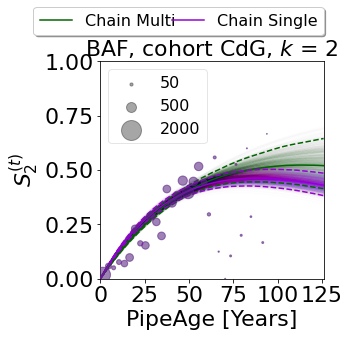}
\caption{}
\end{subfigure}
\caption{Comparisons between different cohorts and chains, for various degradation states and damage types. Find all the figures at \href{https://zenodo.org/record/6535854}{zenodo.org/record/6535854}.}
\label{fig:projections_DTMCs}
\end{figure*}

Through visual inspection conducted in different sections along the sewer pipe length and based on the European standard EN 13508 \citep{EN13508Part1, EN13508Part2}, the damage codes and class present in the sewer pipe (if any) are set. 

A total of 29'667 inspections are registered for 25'507 sewer pipes. We observe that about 88\% of the inspections were carried out between 2005 and 2016. About 87\% of the pipes were inspected at most once, 12\% twice, and less than 1\% three or more times.

As recommended by domain experts, we focus our attention on the damage codes (d.c.): infiltration (BBF), surface damage (BAF), and crack (BAB), which were observed respectively in 44\%, 35\% and 18\% of the inspections.

\subsection{Results}\label{sec:results}

We compare the cohorts and chains visually. For this, we use the results in Fig. \ref{fig:projections_DTMCs}. In order to build these figures, we train a thousand DTMCs using repeated half-sampled bootstrap (Section \ref{sec:DTMC_parameter_inference}), this with the aim to account for uncertainty. Fig. \ref{fig:projections_DTMCs} presents a few of our results, and the figures for all the analysis can be found in Zenodo\footnote[2]{\foottt \zenodo \label{refnote}}.

Fig. \ref{fig:projections_DTMCs}.(a)-(d) present the probability of being in the state $k$ given $\text{PipeAge}$, cohort, chain and damage code. The markers correspond to the discretized table $\hat{S}_k^{(\hat{n})}$ (with $\Delta t = 3 \text{  year}$) and their size visualizes the counts. The dashed lines indicate a 95th percentile \textit{confidence interval} calculated from the projections of the thousand calibrated DTMCs, and the solid line is the median value. The figures comparing cohorts and chains can be found in  Zenodo\footref{refnote} in the folders: \textit{/comparing\_cohorts} and \textit{/comparing\_chains}, respectively. 

Fig.\ref{fig:projections_DTMCs}.(e)-(g) presents the expectations computed with Eq. \ref{eq:expectation}, which corresponds to the expected (average) severity class for a given damage class at a certain PipeAge, the dashed and solid lines are the confidence interval and the median value, respectively. We use these results to ease the comparison between cohorts and can be found in Zenodo\footref{refnote}: \textit{/comparing\_expectations}.
\vspace{-2pt}
\subsubsection{Comparing Cohorts CMW and CR}
Fig. \ref{fig:projections_DTMCs}.(a)-(b) show that older concrete pipes (25 - 75 years) carrying Mixed and Waste content (Cohort CMW) have a higher probability of being at more severe \textit{surface damage} levels (BAF) than concrete pipes carrying Rainwater (Cohort CR). We can also see the same behavior in Fig. \ref{fig:projections_DTMCs}.(e), where the expected severity class of Cohort CMW is higher than Cohort CR for any PipeAge, where the maximum expected class in a 125-years time horizon is $k=3$.

For \textit{cracks} (BAB), we observe complex changes in the degradation pattern that were not properly captured by the DTMCs (e.g., Cohort CMW in Fig. \ref{fig:projections_DTMCs}.(c)). In addition, we find that it is unlikely to find cracks in states $k=2,3,4$\footref{refnote}.

For \textit{infiltration} (BBF), Fig. \ref{fig:projections_DTMCs}.(d) shows that there is an initial probability of roughly 25\% that pipes in Cohorts CMW and CR experience at least some mild infiltrations (i.e., $k>1$).

\subsubsection{Comparing Cohorts PMW and PR}
Fig. \ref{fig:projections_DTMCs}.(f) shows that Cohorts PMW (PVC pipes carrying Mixed and Waste content) and PR (PVC pipes carrying Rainwater) present a similar pattern under \textit{surface damage} (BAF), with a maximum expected severity in a 125-years time horizon of $k<2$. Under the condition of cracks and infiltration, we did not find a significant difference between cohorts\footref{refnote}. 

\subsubsection{Comparing Cohorts CdL and CdG}
Fig. \ref{fig:projections_DTMCs}.(g) compares Cohorts CdL (Concrete pipes and Width$<$500 mm) and CdG (Concrete pipes and Width$\geq$500 mm), and shows that narrow concrete pipes appear to have more severe \textit{surface damage} (BAF) than the wider pipes. 
The same conclusion holds under the condition of \textit{cracks} (BAB)\footref{refnote}. Regarding \textit{infiltration} (BBF), wider pipes seem faster to reach the probabilities of being in a more severe state, however, there is large uncertainty\footref{refnote}.

\subsubsection{Comparing Chains `Multi' and `Single'}\label{sec:comparing_chains}
When comparing Chains `Multi' and `Single', for most of the cases, we did not observe a significant difference between the projections made with one or the other. However, we observe for a few cases e.g., Cohort CdG, surface damage (BAF), in Fig. \ref{fig:projections_DTMCs}.(h), where Chain `Single' shows to transit faster to more severe states compared to Chain `Multi' when PipeAge $>$ 60 years (find the associated values of $P$ in Fig. \ref{fig:markov_chains}, we used $\Delta t = 3 \text{ years}$).

We suspect this is because the chain `Multi' has the possibility of converging in the diagonal values of $P$ to values (very) close to $1$ (e.g., Fig. \ref{fig:markov_chains}.(a), $p_{3,3} = p_{4,4} = 0.9999$), making these states (almost) adsorbing, something that was not observed for Chain `Single' (Fig. \ref{fig:markov_chains}.(b)).

\section{Discussion and Conclusions}\label{sec:discussion_conclusion}

We model sewer pipe degradation in a large-scale case study in the city of Breda by means of discrete-time Markov chains (DTMCs). We describe a methodology to calibrate DTMCs, and visually\footnote[2]{\foottt \zenodo \label{refnote}} compare degradation patterns across cohorts (i.e., groups of sewer pipes sharing similar characteristics) for three types of damage, namely infiltration, surface damage, and cracks.

We find our DTMCs useful for projecting and estimating future degradation states of sewer pipes and comparing cohorts, for example, across expected severity classes. Using this method we conclude, for example, that concrete pipes carrying Mixed and Waste content degrade faster than those carrying Rainwater, which is a phenomenon typically identified in practice.

Comparing the DTMC types, we find that the chains `Multi' and `Single' have similar performance. Although the chain `Single' is simpler, it can be more easily calibrated because it has fewer parameters, and is suitable for this case study. As for the `Multi' chain, it requires a better implementation to avoid the formation of absorbing intermediate states.

In terms of limitations and future directions, when computing the discretized table, we assume that the condition of the sewer pipes is observed at the time of inspection, which may not be the case due to right-censorship (i.e., damage with certain severity happens before the inspection). 
This makes our DTMCs biased, predicting failures later than they actually occur. To address this, we plan to explore \textit{multi-state survival models} for \textit{interval-censored data} \citep{van2016multi}, which leverage survival analysis and better account for censored data. 

Additionally, we will improve the way the parameters of the DTMCs are inferred by accounting for the covariates via \textit{Maximum Likelihood Estimation}. In this way, we will not need to discretize our data based on time intervals.

\newpage
\medskip
\noindent
{\bf Acknowledgement.}
This research has been partially funded by NWO under the grant PrimaVera (https://primavera-project.com) number NWA.1160.18.238, and has received funding from the European Union’s Horizon 2020 research and innovation programme under the Marie Sklodowska-Curie grant agreement No 101008233.

\renewcommand\theequation{A.\arabic{equation}}
\setcounter{equation}{0}


\bibliographystyle{chicago}
\bibliography{refs}
\end{document}